\documentclass[twocolumn,twoside,letterpaper,11pt]{article}
\usepackage{graphicx,xrrc,natbib}
\usepackage{times}

\begin{document}


\title{RADIO DETECTIONS OF ULTRALUMINOUS X-RAY SOURCES}


%
%
%
%


\author{    N. A. Miller                    } 
\institute{ Johns Hopkins University\footnote{Jansky Fellow, National Radio Astronomy Observatory. The National Radio Astronomy Observatory is a facility of the National Science Foundation operated under cooperative agreement by Associated Universities Inc.}        } 
\address{   Department of Physics and Astronomy, 3400 N. Charles Street }
\address{   Baltimore, MD 21218             } 
\email{     nmiller@pha.jhu.edu        } 

\author{ S. G. Neff, R. F. Mushotzky  }
\email{     neff@stars.gsfc.nasa.gov, mushotzky@lheavx.gsfc.nasa.gov}


\maketitle

\abstract{We present some results from an archival {\it VLA} study of ultraluminous X-ray sources (ULXs). These unresolved non-nuclear X-ray sources have luminosities ($L_X \geq 10^{39}$ ergs s$^{-1}$) which may require somewhat exotic explanations, such as intermediate mass black holes or super-Eddington accretion. Radio emission is a powerful way to investigate such sources, through radio morphology and the implications made by source energy and lifetimes derived from the radio. The three galaxies we present here suggest that no single model explains all ULXs, yet there is growing evidence that some ULXs are powered by intermediate mass black holes.
 }

\section{Introduction}

Sensitive, high-resolution X-ray observatories such as {\it Chandra} are identifying numerous unusually strong X-ray sources in nearby galaxies. These sources have X-ray luminosities in excess of X-ray binaries (which have $L_X < \sim10^{38}$ ergs s$^{-1}$), yet fall short of active galactic nuclei (which have $L_X > \sim10^{42}$ ergs s$^{-1}$). What is most intriguing about this luminosity range is that it is above the Eddington limit for spherical accretion onto the end products of normal stellar evolution, yet such sources often do not lie near the nuclei of their host galaxies. Consequently, they are generally called ``intermediate X-ray objects'' (IXO) or more frequently ``ultraluminous X-ray sources'' (ULX). They are generally defined as non-nuclear sources with $L_X \geq 10^{39}$ ergs s$^{-1}$.

What are these enigmatic sources? Multiple theories exist, including the strong possibility that no single model can explain all ULXs. First, ULXs may simply be chance superpositions of sources not physically associated with the apparent host galaxies (mainly background AGN). However, even in the days of {\it ROSAT} it was realized that there are simply too many candidate ULXs to be explained by chance superpositions of background sources \cite{fabbiano89}. Another less exotic potential progenitor is strong young supernova remnants (SNR), which can reach $L_X \sim 10^{40}$ ergs s$^{-1}$. The main weakness of this explanation is that the X-ray emission from ULXs is generally too variable to be explained as one or more SNR. In fact, the X-ray spectra of ULXs are usually consistent with X-ray binaries (XRBs), including those with a black hole primary \citep[e.g.,][]{makishima00}. The question then becomes ``What is the mass of the black hole?'' The importance of this question lies in the distinction between stellar mass black holes and intermediate mass black holes \citep[IMBHs, of $10^2 - 10^4$ M$_\odot$, e.g.][]{colbert99}. ULXs might arise from the former should their emission be beamed, either mechanically \citep[i.e., anisotropic emission due to the accretion disk;][]{king01} or relativistically \citep[akin to microquasars;][]{kording02}. This beaming would alleviate any concern over super-Eddington luminosities \citep[additionally, models with super-Eddington emission from thin accetion disks have been proposed;][]{begelman02}. Obviously, should the accreting object be an IMBH there is no violation of the Eddington limit and the sources can radiate isotropically. This model is attractive as it could fill in the black hole mass function between the stellar mass black holes which arise from normal stellar evolution and the supermassive black holes black holes found at the centers of galaxies. The tradeoff is that we would then require an explanation for how such compact objects formed, although models do indicate they might form via mergers in compact stellar clusters \citep{miller02}.

Radio observations represent a powerful yet often overlooked tool toward evaluating ULXs. The long wavelengths of radio emission are essentially unaffected by the dust extinction which can complicate optical searches, and there are numerous baseline expectations for the relative proportions of radio and X-ray flux for known objects \citep[e.g.,][]{neff03}. At a very simple level, the radio morphology of any detected ULX strongly discriminates between beamed stellar mass XRBs and isotropic emission from an IMBH: unresolved radio emission is consistent with beaming towards the observer, whereas resolved emission suggests any beaming is not along the line of sight (assuming the radio and X-ray emission originate from the same source). Along these lines, \citet{kaaret03} used an unresolved ATCA detection of the ULX in NGC~5408 to argue for a beamed progenitor. Lastly, radio fluxes and spectral indices can be used to evaluate the emission mechanism (e.g., bremsstrahlung, synchrotron) and source total energy and lifetime. These are important clues toward understanding the nature of ULXs.

We are involved in a radio study of known ULXs which relies on archival {\it VLA} data. In particular, we rely on archival data which combines fairly lengthy integration times with higher resolution arrays (e.g., A array at 4.86~GHz). The study is largely blind in that we produce radio images of galaxies hosting ULXs and then compare these with the locations of X-ray point sources identified by {\it Chandra} and {\it XMM}. To date, our findings include a number of nuclear radio and X-ray detections (presumably low luminosity AGN) and several non-nuclear detections. We present here the results for three galaxies which exemplify the diversity of intrinsic non-nuclear source types: Holmberg~II, NGC~3877, and NGC~4631.

\section{Results and Images}

The archive included fairly deep 4.86~GHz imaging for each galaxy, with Holmberg~II and NGC~3877 also possessing useful 1.4~GHz data and NGC~4631 including a deep 8.46~GHz observation. Fluxes and positional centers for detected radio emission were evaluated by fitting Gaussians. All correspondences had separations of radio and X-ray positions of less than 1$^{\prime\prime}$ (although see NGC~4631). Details for each galaxy follow.

\subsection{Holmberg~II}

The radio emission associated with the ULX in Holmberg~II is extended, with a deconvolved size of $\sim3^{\prime\prime} \times \sim2^{\prime\prime}$. The fluxes are $S_{1.4GHz} = 0.98 \pm 0.23$ mJy and $S_{4.86GHz} = 0.70 \pm 0.19$ mJy, translating to a spectral index of $\alpha = -0.27 \pm 0.36$ (under the assumption that the source has not varied over the time of the radio observations, and where $S_\nu \propto \nu ^\alpha$). The implied source lifetime for either thermal bremsstrahlung or synchrotron (under equipartition) is of order $10^8$ years. The ratio of radio to X-ray flux, $R_X \equiv L_\nu \nu / L_X$ \citep[where $\nu = 5$ GHz; see][]{neff03}, is about $6 \times 10^{-6}$, much lower than that of starforming regions and supernova remnants and consistent with sources such as microquasars.

\begin{figure}
  \begin{center}
    \includegraphics[width=\columnwidth, angle=270]{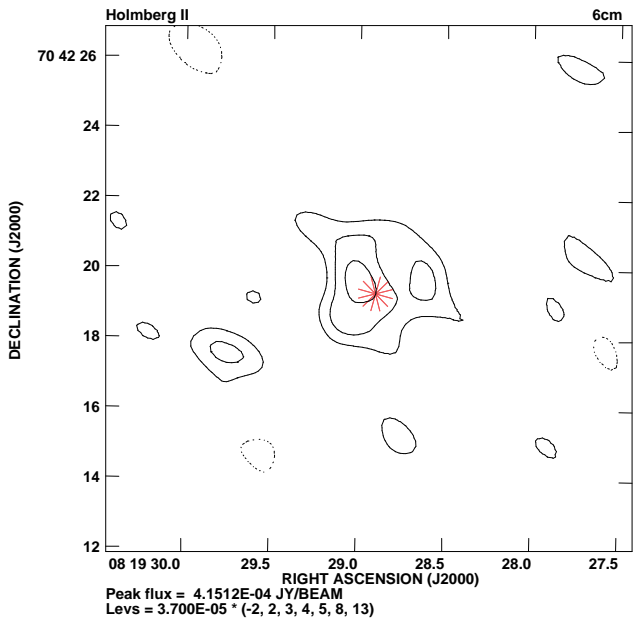}
    \caption{\small Radio emission at 4.86~GHz for the ULX in Holmberg~II. The location of the X-ray source is indicated by the red asterisk, whose size corresponds to the 1$\sigma$ positional error.}
    \label{fig-ho2}
  \end{center}
\end{figure}

It is interesting to note that Holmberg~II is associated with an optical nebula \citep{pakull03}. In fact, the radio emission covers an area of the same physical size as the ionized gas ($\sim35$ pc). \citet{pakull03} use nebular models and the detected luminosity for the He {\scshape ii} $\lambda$4686 $\mbox{\AA}$ line to deduce an isotropically-emitting source whose X-rays ionize the surrounding nebula. The ULX in Holmberg~II is also among the better candidates for a true black hole binary on the basis of its well-observed X-ray spectrum and variability \citep{dewangan04}.

\subsection{NGC~3877}

In the case of NGC~3877, there are two X-ray point sources associated with radio emission. One is well known, as it is SN1998S. The other is quite close to the nucleus, yet the mutually consistent radio and X-ray position is offset by several arcseconds from the optical nucleus in an {\it HST} image. The radio emission from SN1998S is unresolved, with $S_{4.86GHz} = 1.12 \pm 0.10$ mJy yielding an $R_X$ of about $10^{-4}$. The near-nuclear source is resolved, with a size of $\sim4^{\prime\prime} \times \sim2^{\prime\prime}$. Its radio fluxes are $S_{1.4GHz} = 2.64 \pm 0.41$ and $S_{4.86GHz} = 2.42 \pm 0.55$, for a spectral index of $\alpha = -0.07 \pm 0.28$. The corresponding $R_X$ is about $5 \times 10^{-3}$, and again we find cooling times of $\sim 10^8$ years.

\begin{figure}
  \begin{center}
    \includegraphics[width=\columnwidth, angle=270]{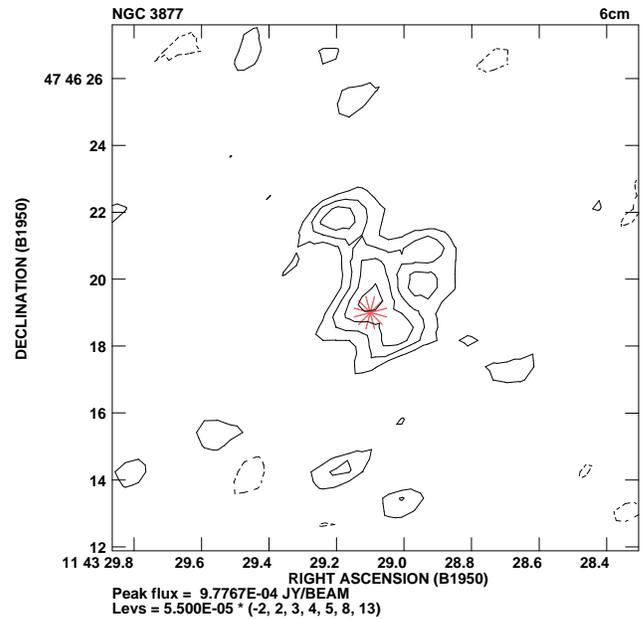}
    \caption{\small Radio emission at 4.86~GHz for the near-nuclear ULX in NGC~3877.}
    \label{fig-n3877}
  \end{center}
\end{figure}

\subsection{NGC~4631}

The radio emission associated with an X-ray point source in NGC~4631 is resolved into four aligned components. The morphology is suggestive of a background radio galaxy, and its radio emission is consistent with this interpretation. The net fluxes are $S_{4.86GHz} = 4.58 \pm 0.22$ mJy and $S_{8.46GHz} = 3.60 \pm 0.11$ mJy ($\alpha = -0.43 \pm 0.06$), yielding an $R_X$ of about $2 \times 10^{-3}$. Its X-ray luminosity of $8 \times 10^{37}$ ergs s$^{-1}$ (calculated assuming it does lie in NGC~4631) is also below the formal cutoff for ULXs.

\begin{figure}
  \begin{center}
    \includegraphics[width=\columnwidth]{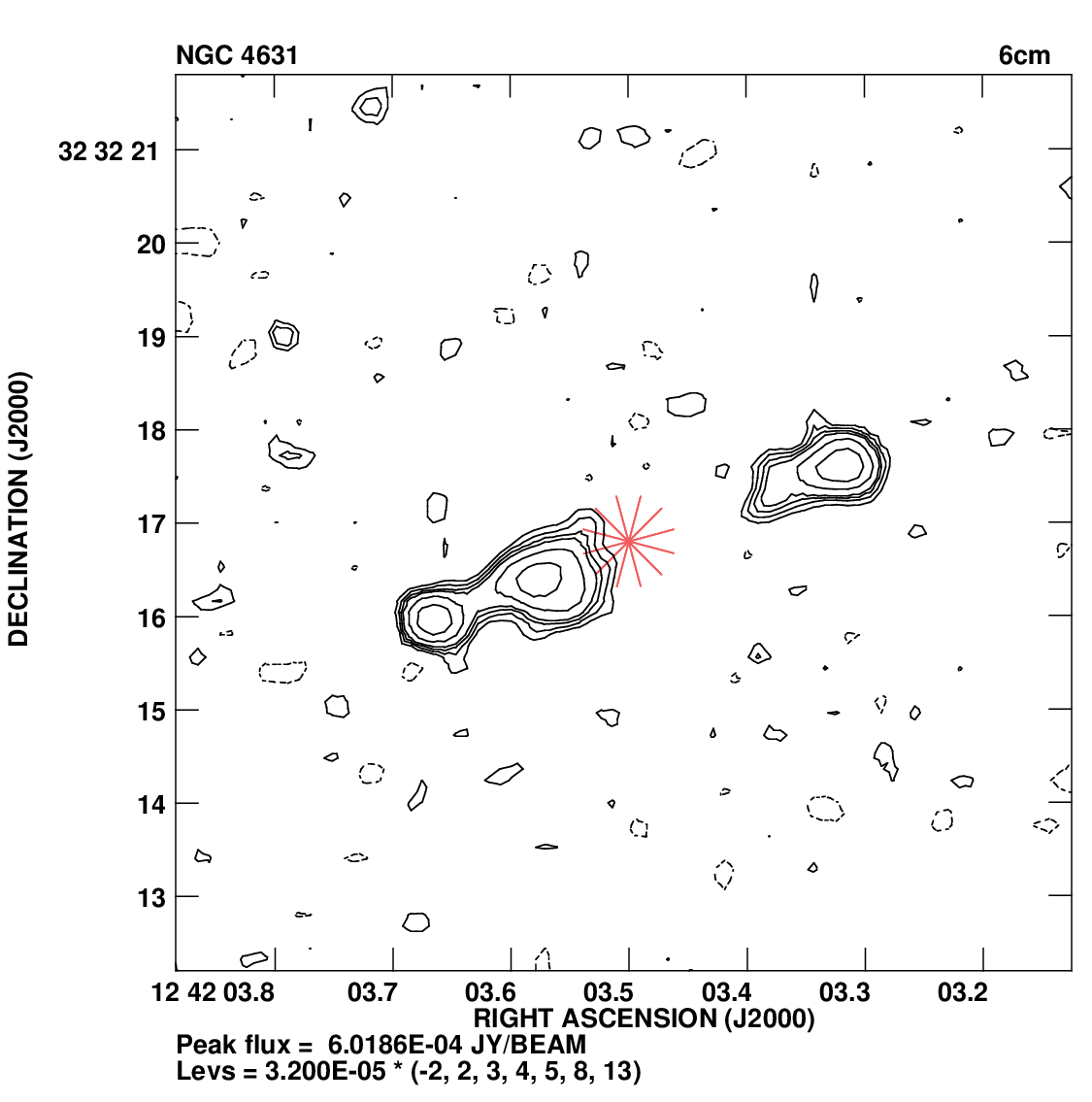}
    \caption{\small Radio emission at 4.86~GHz for the X-ray source in NGC~4631.}
    \label{fig-n4631}
  \end{center}
\end{figure}

\section{Discussion}

The results for these three galaxies are illustrative of some of the difficulties in understanding ULXs. In general, no simple picture can fit all four sources and this suggests that ULXs are a mixed bag of source types. The ULX in Holmberg~II is quite enticing, with extended radio emission arguing against beamed progenitors and suggesting IMBHs as the root of ULXs. However, the ULX in NGC~5408 has a comparable $R_X$ value yet is apparently compact based on the ATCA radio data described in \citet{kaaret03}. We also have one clear SNR (SN1998S in NGC~3877), with the other source in NGC~3877 presenting some confusion. Although this second source is offset from the optical nucleus by a few arcseconds, its $R_X$ places it between expectations for low luminosity AGN and other radio-detected ULXs. Thus, it may represent a low luminosity AGN \citep[similar to the pair of low luminosity AGN in the merger NGC~3256;][]{neff03} or a ULX similar to the one in Holmberg~II but whose X-ray observation happened to occur at a time of unusually low luminosity. Lastly, the radio-detected X-ray source in NGC~4631 appears to be a background AGN.

Finally, we stress that the variety of intrinsic source types underscores the utility of radio observations. The radio data greatly aid in the characterization of the candidate ULXs, and continuation of such studies should lead to significant numbers of the various source types and ultimately to an answer to the question of what powers these mysterious objects.

\section*{Acknowledgments}

We thank the organizers of this conference for an entertaining and well-run meeting, and those {\it VLA} observers whose archival data we have been fortunate enough to use.

\end{document}